\documentclass[3p]{elsarticle}
\usepackage{amssymb}

\usepackage{graphicx}
\usepackage{balance}
\usepackage{color}
\usepackage{xcolor}
\usepackage{multirow}
\usepackage{caption}
\usepackage{subcaption}
\usepackage[utf8]{inputenc}
\usepackage{amsfonts}
\usepackage{amsmath}
\usepackage{graphicx}
\usepackage{float}
\usepackage[ruled,vlined]{algorithm2e}
\DeclareMathOperator*{\argmin}{arg\,min}

\journal{Smart Health}

\begin{document}
\begin{frontmatter}

\title{ Inter-Mobile-Device Distance Estimation using Network Localization Algorithms for Digital Contact Logging Applications}

\author[usc]{Lillian Clark\corref{cor1}}
\ead{lilliamc@usc.edu}
\author[mit]{Alan Papalia}
\ead{apapalia@mit.edu}
\author[ufsc]{J\^{o}nata Tyska Carvalho}
\ead{jonata.tyska@ufsc.br}
\author[cca]{Luca Mastrostefano}
\ead{luca.mastrostefano90@gmail.com}
\author[usc]{Bhaskar Krishnamachari}
\ead{bkrishna@usc.edu}
\address[usc]{Ming Hsieh Department of Electrical and Computer Engineering, University of Southern California, United States}
\address[mit]{Computer Science and Artificial Intelligence Laboratory, Massachusetts Institute of Technology, United States}
\address[ufsc]{Department of Informatics and Statistics (INE), Federal University of Santa Catarina, Brazil}
\address[cca]{Covid Community Alert, United Kingdom}
\cortext[cor1]{Corresponding author}

\begin{abstract}
    Mobile applications are being developed for automated logging of contacts via Bluetooth to help scale up digital contact tracing efforts in the context of the ongoing COVID-19 pandemic. A useful component of such applications is inter-device distance estimation, which can be formulated as a network localization problem. We survey several approaches and evaluate the performance of each on real and simulated Bluetooth Low Energy (BLE) measurement datasets with respect to both distance estimate accuracy and the proximity detection problem. 
    We investigate the effects of obstructions like pockets, differences between device models, and the environment (i.e. indoors or outdoors) on performance.
    We conclude that while direct estimation can provide the best proximity detection when Received Signal Strength Indicator (RSSI) measurements are available, network localization algorithms like Isomap, Local Linear Embedding, and the spring model outperform direct estimation in the presence of missing or very noisy measurements. The spring model consistently achieves the best distance estimation accuracy.
\end{abstract}

\end{frontmatter}

\section{Introduction}

Mobile apps for automated contact logging using Bluetooth are being developed by many organizations to help contain the ongoing COVID-19 pandemic \cite{covidcommunityalert,pact,bay2020bluetrace}. 
While many of these apps only detect if two devices are in proximity, it is helpful to also estimate the distance between the devices more accurately in order to determine the risk of possible infection. Distance can be estimated via the received signal strength of a Bluetooth beacon \cite{leith2020coronavirus}. 
As the distance increases, the signal received weakens. However, these signals are often susceptible to high levels of statistical fluctuations due to scattering, absorption and reflections in the environment, leading to high variance in the signal strength measurements and low confidence in the distance estimate. This uncertainty has lead many in the community, including Sven Mattison who co-invented Bluetooth, to doubt the feasibility of contact logging apps which rely on ``vanilla point-to-point Bluetooth links" \cite{6487509}.

\begin{figure}
    \centering
    \includegraphics[width=0.5\textwidth]{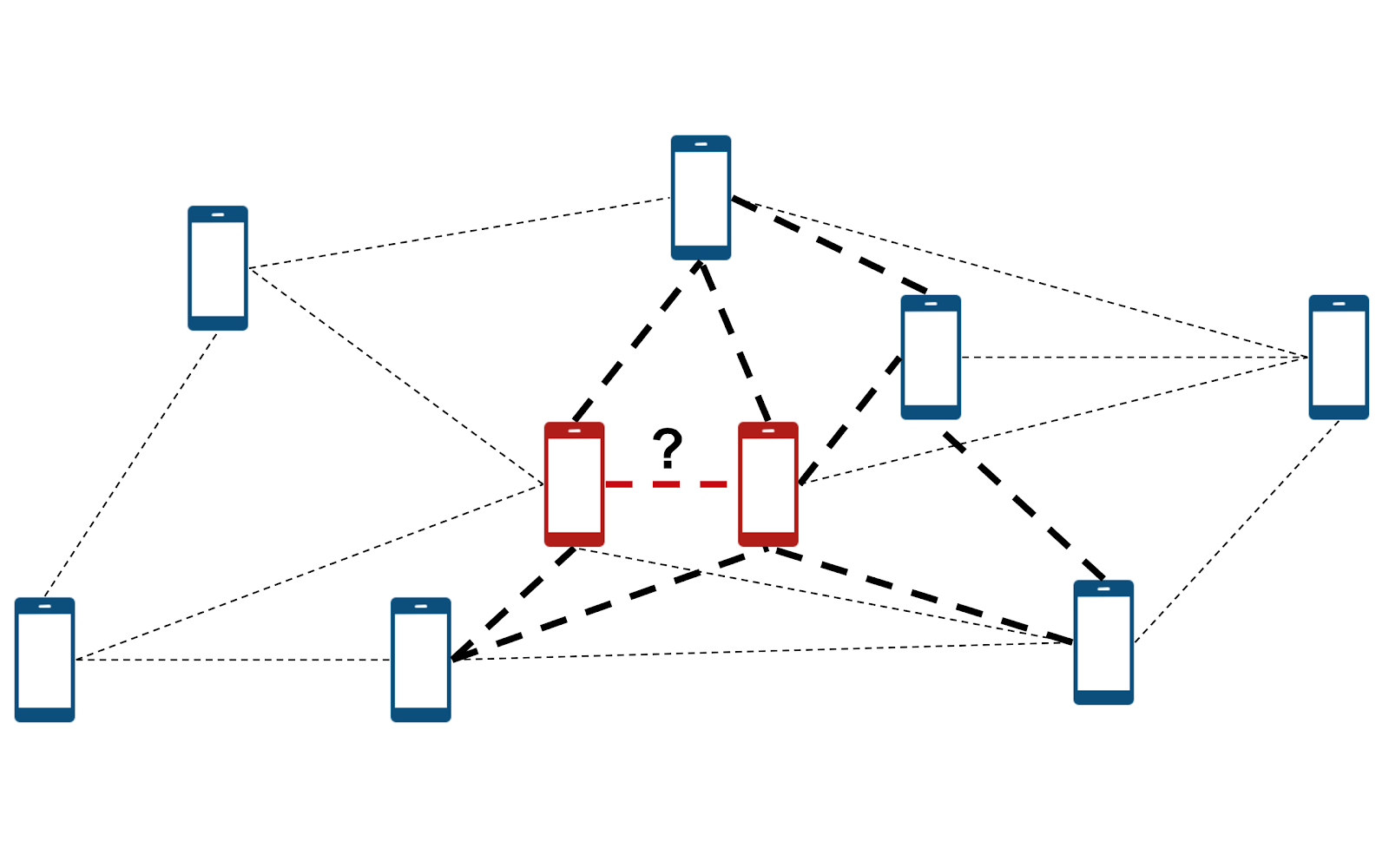}
    \caption{A network of available RSSI measurements, with more reliable measurements bolded. Our goal is to leverage all available measurements to get accurate distance measurements for any pair of devices, and improve accuracy for pairs with available measurements like the one shown in red.}
    \label{fig:schematic}
\end{figure}

To overcome this, we can make use of the fact that in populated areas there are typically more than two devices present. For example, on buses, in grocery store lines, and in public parks, connectivity can be modeled with a graph, as in Figure \ref{fig:schematic}. Edges on the graph represent the presence of Received Signal Strength Indication (RSSI) measurements between two devices. Shorter edges, bolded for emphasis, represent stronger signals which are typically less noisy and lead to better estimates. While we may have an RSSI-based distance estimate between two devices, like those shown in red, our objective is to improve on this estimate by leveraging the other available RSSI measurements. We can also use available RSSI measurements to estimate the distance between two devices for which a measurement is unavailable.
This is possible assuming a centralized architecture in which RSSI measurements are stored in a central server. While decentralized solutions have also been proposed, they pose various challenges \cite{bay2020bluetrace}. The availability of a central server allows us to explore methods which improve distance estimation accuracy and will allow us to more confidently judge whether a device owner has been at risk.

The contribution of this work is a quantitative comparison of algorithms from the sensor network and learning communities which perform localization and can be used to estimate inter-device distances. In considering these methods, we focus on accuracy in distance estimation and proximity detection, i.e. detecting if two devices are within two meters. We evaluate each method on real and simulated data, with special attention to noisy and missing RSSI measurements, scalability, computational complexity, and common contact logging use cases. We qualitatively summarize our findings in Table \ref{table:key_findings} and present more detailed summaries in Section \ref{sec:summary}.


\begin{table}[h!]
  \begin{center}
    \caption{Key Findings}
    \label{table:key_findings}
    \begin{tabular}{l|l|l}
      Approach & Advantages & Disadvantages \\
      \hline\hline
    Direct Estimation & Best proximity detection & Not robust to missing RSSIs \\
    & Lowest complexity & \\
    \hline
    Semidefinite Programming & Robust to missing RSSIs & Worst false positive rate \\
    & Best true positive rate & \\
    & Good distance estimation & \\
    \hline
    Spring Model & Robust to missing RSSIs & Some variance in performance \\
    & Best distance estimation & Potential for divergent solutions\\
    & Good proximity detection & \\
    \hline
    Non-linear manifold learning & Good proximity detection & \\
     & Good distance estimation & \\
    \hline
    Multidimensional Scaling & & Not robust to missing RSSIs \\
    (metric) & & \\
    \hline
    Multidimensional Scaling & Good distance estimation & Poor proximity detection \\
    (non-metric) & & \\
    \end{tabular}
  \end{center}
\end{table}

This paper is organized into \ref{sec:conclusion} sections. Section \ref{sec:background} presents relevant preliminaries including background, related work, and the RSSI-to-distance model used. Section \ref{sec:problem} concisely formulates the problem. Section \ref{sec:approaches} presents the approaches evaluated in this work, with the appropriate details for reproducibility. Section \ref{sec:experiment} presents the results on data collected from real devices and Section \ref{sec:simulated} presents experiments on simulated data to augment these results. We conclude with a summary and discussion of future work in Section \ref{sec:conclusion}. All relevant code for this paper can be found in \cite{github_repo}.



\section{Background and Related Work}
\label{sec:background}

In this work we are interested in determining the Euclidean distances between $n$ devices or \textit{nodes} based on measurements between them which are closely related to distance. 
To find the complete Euclidean distance matrix, we can first solve the problem of localizing $n$ nodes in two-dimensional space. Note that since we are only interested in the relative locations, translations and rotations of the $n$ points do not matter.
The problem of mapping $n$ points to two-dimensional space (or more generally, to a low-dimensional space) based on distances (often called \textit{dissimilarities}) between them is well-studied and has a variety of applications.

In information visualization, the related problem of displaying complex data in 2D is commonly called graph realization, graph drawing, or graph embedding \cite{8294302}. 
In cyberphysical systems, determining the physical locations of devices is typically called sensor network localization \cite{bachrach2005localization}. 
In machine learning, embedding datasets which exist in high-dimensional space into low-dimensional manifolds can be highly useful for things like semantic analysis, and is commonly called manifold learning \cite{izenman2012introduction}.

Many approaches exist for solving the network localization problem. Multidimensional scaling (MDS) is a technique originally developed for use in mathematical psychology, and falls into two categories: metric or non-metric. 
In metric MDS, the input dissimilarity matrix is assumed to behave like distance, i.e. dissimilarities are a function of an underlying metric and is typically used whenever quantitative data is available \cite{young1985multidimensional,dissimilarity}.
The goal of a metric MDS algorithm is to find a configuration of points in space such that the distances between two points are as close as possible to the dissimilarity data \cite{manifold}.
A non-metric MDS algorithm, by contrast, will try to preserve the order of the distances by finding a monotonic relationship between the distances in the two-dimensional space and the dissimilarities. 
Non-metric MDS is typically used when only qualitative data is available \cite{young1985multidimensional}, although it is straightforward to derive ordinal data when quantitative data is available.

The goal of metric MDS can be written as an optimization over all possible configurations which minimizes a loss function. \textit{Stress} is a commonly used loss function given by
\begin{equation}
    \textrm{Stress}(\mathcal{X}) = \sqrt{\frac{\sum_{i=1}^{n}\sum_{j=1}^{i-1}(d_{ij}-\hat{d}_{ij})^2}{\sum_{i=1}^{n}\sum_{j=1}^{i-1}\hat{d}_{ij}^2}}
\end{equation}
where $d_{ij}$ is the dissimilarity between $i$ and $j$ and $\hat{d}_{ij}$ is the distance between $i$ and $j$ in configuration $\mathcal{X}$.
The classical approach, also called Principal Coordinates Analysis (PCoA), uses eigenvalue decomposition to find the optimal configuration $\mathcal{X}$ \cite{torgerson1952multidimensional}.
An excellent description of classical MDS can be found in Appendix C of \cite{bachrach2005localization}.

The disadvantage of classical MDS is that it requires dissimilarity measurements for \textit{all} pairs of nodes, which are often not available in network localization problems. Work in distance matrix completion using graph distances or shortest paths can be used to first estimate a complete dissimilarity matrix, which allows the use of classical MDS \cite{shang2004localization,shang2004improved}.
Another disadvantage of classical MDS is that it does not perform well when the dissimilarities are nonlinear. Isomap and Locally Linear Embedding (LLE) are two algorithms which extend classical MDS for nonlinear embeddings which use nearest neighbor search and are well-suited for manifold learning \cite{manifold}.

Other approaches to finding the optimal configuration $\mathcal{X}$, sometimes called stress majorization, take an iterative approach. A seminal work from Kruskal in the early 1960s describes an iterative steepest descent approach, wherein each point in an initially random configuration is adjusted slightly along the negative gradient of the stress with respect to that point's position. An extension of that work presents the Scaling by Majorizing a Complicated Function (SMACOF) algorithm, which is the basis for the multi-dimensional scaling function provided by the scikit-learn manifold library \cite{de2005applications}.

The previously mentioned approaches are centralized algorithms, however distributed variations exist. This class of algorithms starts with each node estimating its position, and then refining or relaxing this estimate based on local information i.e. the dissimilarity and position estimates of its neighbors. Local neighborhood multilateration is presented in \cite{savarese2001location}, where each node adjusts its position by using its neighbors as temporary beacons (reference points).
An equivalent formulation to local multilateration is presented in \cite{priyantha2003anchor} and is generally referred to as a spring model. 
In the graph formulation, edges between nodes are imagined as springs where the equilibrium length is the measured dissimilarity. Each node iteratively adjusts its position in the direction of its local spring force. While elegant, spring models are highly sensitive to initial starting positions, which can lead the algorithm to get stuck in local minima. One solution is to use unweighted graph distances to formulate a ``fold-free" initial configuration \cite{priyantha2003anchor}.
Another approach is to relax the problem into a semidefinite program and apply the solution of the program as the initial configuration \cite{Biswas2006approach}.


In the semidefinite program (SDP) approach to network localization, nonconvexities in the problem formulation are replaced with convex components in a way that attempts to closely model the nonconvex form of the original problem, and the resulting convex problem is then solved using common optimization machinery \cite{doherty2001convex, Biswas2006approach}.
Solutions to the SDP can be solved with approaches like the interior point algorithm, and these solutions can then be further refined by use of local, gradient-based optimization methods with the original problem formulation.

\subsection{Related Work}

Several surveys on localization exist in the sensor network community. The authors of \cite{mao2007wireless} present an overview of measurement techniques, one-hop localization algorithms, and multi-hop localization algorithms including MDS and SDP, with a focus on the discussion of centralized versus distributed algorithms. Wang \emph{et al.}~\cite{wang2010survey} present a taxonomy of localization algorithms and performance analysis with respect to location accuracy, density and portion of the network with \textit{a priori} known positions, computation cost and communication cost, while Chowdhury \emph{et al.}~\cite{chowdhury2016advances} follow up with advances and recent trends. Our work, by contrast, focuses on different metrics including node-node distance estimation and proximity detection. In \cite{langendoen2003distributed}, simulation results are presented for three localization algorithms with a useful discussion of inter-node distance error. The algorithms presented depend on trilateration and the presence of three or more anchors (nodes with known initial positions), which is an assumption not made in this work. They conclude that no option performs the best under all conditions, motivating our investigation of network localization algorithms in this application-focused context.

The authors of \cite{yang2006distance} present a survey of distance metric learning, highlighting supervised methods and unsupervised methods like Isomap and LLE. Saeed \emph{et al.}~\cite{saeed2018survey} present an overview of multidimensional scaling and its extensions, with specific attention to different loss functions. In \cite{van2009dimensionality}, the authors present the results of linear and non-linear dimensionality reduction algorithms on real and simulated datasets. They are interested in the metrics of nearest-neighbor classification, continuity, and trustworthiness of the embedding, while this paper is specifically concerned with distance estimation accuracy and proximity detection. The authors of \cite{patwari2004manifold,kashniyal2017wireless} demonstrate the feasibility of using manifold learning algorithms for sensor network localization, bridging these related research thrusts and providing further motivation for this work.

Research in proximity detection for contact logging via RSSI measurements has been primarily concerned with feasibility and the relationship between RSSI and distance \cite{leith2020coronavirus,leith2020measurement}, with a parallel research thrust in privacy-preservation \cite{cho2020contact,li2020covid}. The contribution of this work is a survey of network localization algorithms and a comparative evaluation of their performance with respect to inter-mobile-device distance estimation and proximity detection in real and simulated data.

\subsection{Pathloss Model}

To estimate distance from RSSI, we use a radio propagation model which predicts the path loss a signal encounters over distance. In this work we use a simple path loss model with log-normal fading, a common noise model for RSSI signals. Formally,
\begin{equation}
    \label{eq:pr}
    P_r = P_T K \left(\frac{d}{d_0}\right)^\eta \psi
\end{equation}
where $P_r$ is the received power, $P_T$ is the transmitted power, $d$ is the distance between the devices, $d_0$ is the reference distance, $K$ is the gain at the reference distance (a number smaller than 1). The pathloss exponent $\eta$ capture how quickly the signal falls off with distance, and is typically between 2 and 3. The noise is captured in the random variable $\psi$. Power is commonly measured on a decibel scale, where equation (\ref{eq:pr}) is equivalently
\begin{equation}
    P_{r_{dBm}} = P_{T_{dBm}} + K_{dB} - \eta 10 \log_{10} \frac{d}{d_0} + \psi_{dB}
\end{equation}
where $\psi_{dB}$ is now a zero-mean Gaussian random variable with variance $\sigma_{dB}^2$.
It is often easier to assume the mean power received at a reference distance, $P_{0_{dBm}}$ is known, and therefore
\begin{equation}
    P_{r_{dBm}} = P_{0_{dBm}} - \eta 10 \log_{10} \frac{d}{d_0} + \psi_{dB}
\end{equation}
Note that $P_{r_{dBm}}$ is the measured RSSI, and it decreases linearly with log-distance. This means that the parameters $P_{0_{dBm}}$, $d_0$, and $\eta$ can be fitted using linear regression if a sample dataset is available. The standard deviation $\sigma_{dB}$ can also be calculated from available data.

To calculate the distance estimate, $d$, if these parameters are known, we simply rearrange terms,
\begin{equation}
    \label{eq:distance_from_RSS}
    d = d_0 (10^{-\frac{(P_{r_{dBm}} - P_{0_{dBm}})}{10\eta}} )
\end{equation}
\begin{equation}
\label{eq:dmin}
    d_{min} = d_0 (10^{-\frac{(P_{r_{dBm}} - P_{0_{dBm}} + 2\sigma_{dB})}{10\eta}} )
\end{equation}
\begin{equation}
\label{eq:dmax}
    d_{max} = d_0 (10^{-\frac{(P_{r_{dBm}} - P_{0_{dBm}} - 2\sigma_{dB})}{10\eta}} )
\end{equation}
where distance is between $d_{min}$ and $d_{max}$ with 95.45\% confidence.

\section{Problem Formulation}
\label{sec:problem}
The objective is to accurately estimate inter-device distances based on a matrix of available RSSI measurements. A possible matrix $\mathbf{RSSI}$ is shown below,
$$
\begin{bmatrix} 
0 & -65 & \emptyset & -68 & -66\\
-66 & 0 & -72 & \emptyset & -73\\
\emptyset & -73 & 0 & -68  & -71\\
-68 & \emptyset & -68 & 0 & -59\\
-66 & -71 & -71 & -58 & 0 \\
\end{bmatrix}
\quad
$$
where the measurement in the $i^{th}$ row and $j^{th}$ column represents the signal received at the $i^{th}$ device and transmitted by the $j^{th}$ device.
The zeros along the diagonal represent the distance between a device and itself. $\emptyset$ represents the absence of a measurement. Notice that for device pair $i$ and $j$, the signal received by $i$ may differ in strength from the signal received by $j$, such that the matrix is not symmetric.

From this matrix we seek to produce a symmetric matrix of distance estimates $\mathbf{\hat{d}}$, where ${\hat{d}_{ij}}$ is the estimated distance between devices $i$ and $j$. For example,
$$
\begin{bmatrix} 
0 & \hat{d}_{12} & \hat{d}_{13} & \hat{d}_{14} & \hat{d}_{15}\\
\hat{d}_{12} & 0 & \hat{d}_{23} & \hat{d}_{24} & \hat{d}_{25}\\
\hat{d}_{13} & \hat{d}_{23} & 0 & \hat{d}_{34} & \hat{d}_{35}\\
\hat{d}_{14} & \hat{d}_{24} & \hat{d}_{34} & 0 & \hat{d}_{45}\\
\hat{d}_{15} & \hat{d}_{25} & \hat{d}_{35} & \hat{d}_{45} & 0\\
\end{bmatrix}
\quad
$$

In some cases we first find the configuration $\mathcal{X} = \{x_1,x_2,...x_n\}$ where $x_1,...,x_n \in \mathcal{R}^{2}$. This represents the location of each node in two-dimensional space, and we will refer to finding $\mathcal{X}$ as the network localization problem. Given $\mathcal{X}$, we can calculate $\mathbf{\hat{d}}$ since $\hat{d}_{ij} = \left\|x_{i}-x_{j}\right\|$.

\section{Approaches to Estimating Inter-Device Distances}
\label{sec:approaches}

\subsection{Direct Estimation}

The simplest approach is to consider each RSSI measurement individually. In this case, for each entry in the \textbf{RSSI} matrix, we have one entry in the corresponding $\hat{\textbf{d}}$ matrix calculated with equation (\ref{eq:distance_from_RSS}). This is the solution provided by direct estimation, but it is also the starting point for all other approaches. We refer to the set of node-node distance measurements available directly from the pathloss model as $\mathcal{N}$.

Note that direct estimation does not produce a symmetric distance matrix. We can choose to average the RSSI measurements prior to calculating distance, which we refer to as the pre-averaged approach, or average the calculated distance measurements, which we refer to as the post-averaged approach. Direct estimation also cannot produce distance estimates for missing RSSI for device pairs without RSSI measurements. We remedy this by assuming a threshold RSSI value (e.g. $-95dB$) for missing measurements, which represents the noise floor of the receiver. This allows us to use the same evaluation metrics across the various approaches, as certain approaches require measurements between all nodes while others can use sparse RSSI matrices.

\subsection{Semidefinite Programming}
Distances can be estimated by performing network localization. To solve the network localization problem, we use the open source convex optimization libraries CVXPY \cite{diamond2016cvxpy, agrawal2018rewriting} and SCS \cite{scs1, scs2}, which use the alternating direction method of multipliers to solve a previously obtained semidefinite program (SDP) \cite{Biswas2006approach}. This approach makes use of what
are known as anchors and nodes. Anchors are sensors in which the location is
known \textit{a priori}.  Nodes are sensors in which
position is unknown, where these locations 
are the estimated variables of interest. Generally, an increased
number of anchors leads to an improved final solution \cite{Biswas2006_gen}, however
anchors are not required for our specific approach \cite{Biswas2006approach}. We present the generalized version of our SDP solution, with the understanding that in the case of inter-device distance estimation there are potentially no anchors. Numerical results suggest that in the case of no anchors, arbitrarily assigning a single node to be an anchor with an
arbitrary, predefined location can improve solver stability and solution quality.

The approach we take looks to solve a problem of the form below, which is
the form of Maximum Likelihood Estimation (MLE) measurements with additive
Gaussian noise \cite{Biswas2006approach, Beck2008}.

\begin{equation} \label{eq:snl_mle}
   \argmin _{\mathcal{X}}\left\{\begin{array}{l}
\sum_{(i, j) \in \mathcal{N}} \frac{1}{\sigma^2_{i j}}\left|\left\|x_{i}-x_{j}\right\|^{2}-d_{i j}^{2}\right| \\
+\sum_{(i, j) \in \mathcal{M}} \frac{1}{\sigma^2_{i j}}\left|\left\|x_{i}-a_{j}\right\|^{2}-d_{i j}^{2}\right|
\end{array}\right\}
\end{equation}

Seen above, $\mathcal{N}$ is the set of all available node-node distance measurements,
$\mathcal{M}$ is the set of all anchor-node distance measurements, $\sigma_{ij}$
is the variance of measurement between sensors $i$ and $j$,
$x_{i}$ and $x_{j}$ are variables representing the positions of nodes $i$ and $j$, $a_{j}$ is a
constant vector representing the known position of anchor $j$, and $d_{ij}$ represents
the distance measurement between the corresponding sensors $i, j$.

However, it is easily shown that the objective function displayed in
equation (\ref{eq:snl_mle}) is non-convex, and as such we cannot guarantee the ability
to optimally solve the problem with standard convex solvers. To remedy this, a convex 
relaxation can be made
which replaces the non-convexities in a manner which attempts to preserve the
general form of the problem. In the implementation chosen, a semidefinite
relaxation is made which relaxes a non-convex quadratic constraint in the
problem. For those interested in the particular details of the formulation, we
direct you to \cite{Biswas2006approach}.

It is notable that though reformulating the problem into this convex form allows
for usage of convex programming and guarantees the ability to find the global
minimum of the relaxed problem, it comes at a loss of accuracy in the cost
function. As a result, though the optimal solution is often a good estimate for
the network localization problem it is oftentimes not the global optimum. For this reason, the
SDP solution is often used as a starting point for a local gradient-based solver
of the original, non-convex problem.

\subsection{Multidimensional Scaling}
To find a solution to the network localization problem, we can use the MDS class provided by the Scikit-Learn Manifold library. This class implements the SMACOF iterative stress majorization algorithm, where at each step of the algorithm a Guttman transform of the current configuration is calculated \cite{guttman1968general}.
A thorough description of the theory, implementation, and convergence results of SMACOF can be found in \cite{de2011multidimensional}.

The MDS class can be configured to perform metric or non-metric MDS. Each takes as input a dissimilarity matrix, and our implementation provides the matrix given by the set of node-node distance measurements $\mathcal{N}$, and seeks to minimize
\begin{equation}
    \textrm{Stress}_{\textrm{SMACOF}}(\mathcal{X}) = \sum_{(i,j)\in\mathcal{N}} \left\|x_{i}-x_{j}\right\| - g(d_{ij})
\end{equation}
where $d_{ij}$ is the node-node distance measurement between $i$ and $j$ and $g(d_{ij})$ is a linear function for metric MDS and monotonic function for non-metric MDS.

Note that the metric configuration requires a complete dissimilarity matrix i.e. the set $\mathcal{N}$ must contain node-node distance measurements for all pairs. We again use the noise floor of the receiver when an RSSI measurement is not available. The non-metric implementation treats $d_{ij}=0$ for $i \neq j$ as a missing entry rather than a distance of zero \cite{manifold}.
The solution $\mathbf{\hat{d}}$ produced by non-metric MDS is then scaled by the average $d_{ij}$ divided by the average $\hat{d}_{ij}$ to produce a final network localization solution.

\subsection{Spring Model}

We present here a spring model (SM) that solves the network localization problem iteratively in a distributed manner in order to minimize the stress on each node,
\begin{equation}
        \textrm{Stress}_{\textrm{SM}}(x_i) = F(x_i) = \sum_{j:(i,j)\in\mathcal{N}} f(x_i,x_j)
\end{equation}
\begin{equation}
\label{eq:springforce}
        f(x_i, x_j) = \frac{\left\|x_{i}-x_{j}\right\| - d_{ij}}{d_{ij,max}-d_{ij,min}} \langle \frac{x_j-x_i}{\left\|x_{i}-x_{j}\right\|} \rangle
\end{equation}
where we refer to stress as $F(x_i)$ as it is the sum of forces applied to node $i$ by each node $j$. These forces, $f(x_i,x_j)$, are analogous to the forces that would be applied if each node-node distance were a stretched or compressed spring. Note that these forces are vectors in the direction of $\langle x_j - x_i \rangle$, and the resulting stress is also a vector. In equation (\ref{eq:springforce}), $d_{ij,max}$ and $d_{ij,min}$ can be calculated from equations (\ref{eq:dmin},\ref{eq:dmax}) and their difference provides a measure of uncertainty.  Scaling the stress by this term allows distances with less uncertainty (typically shorter distances) to have a greater impact on the solution than distances with greater uncertainty.

Our spring model implementation is described in Algorithm \ref{alg:SM}, where $n$ is the number of nodes, $\mathcal{N}$ is the set of node-node distances, $\epsilon$ is a parameter which defines convergence (we use 0.1), and $\gamma$ is a coefficient applied to the force similar to a step size. To prevent the algorithm from diverging, we choose a step size of roughly $\frac{1}{n}$. We can optionally decrease $\gamma$ over time for an adaptive step size, or update $x_i \leftarrow x_i + \gamma (F^t + F^{t-1})$ to include momentum in the gradient descent.

\begin{algorithm}[H]
\SetAlgoLined
    \SetKwInOut{Input}{input}
    \SetKwInOut{Output}{output}
    \Input{$n,\mathcal{N}, \epsilon, \gamma$}
    \Output{$\mathcal{X}$}
    initialize $\mathcal{X}$ randomly in $\mathcal{R}^2$\newline
    \For{\textrm{iteration} t}{
        \For{\textrm{node} $i$}{
            $F$ = 0 \newline
            \For{\textrm{node} $j$}{
                \If{$(i,j) \in \mathcal{N}$}{
                    $F \leftarrow F + f(x_i)$
                }
            }
            $x_i \leftarrow x_i + \gamma F$
        }
        stop if $\left\|x_{i}^t-x_{i}^{t-1}\right\| \leq \epsilon   \forall i \in \{1,...n\}$
    }
 \caption{Spring model for network localization}
 \label{alg:SM}
\end{algorithm}

\subsection{Non-linear Manifold Learning}

Along with the MDS class, the Scikit-Learn Manifold library provides several other classes for generalizing classical MDS to non-linear structures in data.
In this work we implement Isomap to solve the network localization problem \cite{tenenbaum2000global}. Isomap seeks to maintain geodesic distances between all points. Imagining the data as a graph where vertices are nodes and edges are available dissimilarity measures, the crux of Isomap is in estimating the distance between nodes which are not neighbors as the shortest path between them. Then Isomap performs eigenvalue decomposition to solve for node locations.

We also implement Local Linear Embedding (LLE), which preserves dissimilarities within local neighborhoods \cite{roweis2000nonlinear}. The crux of LLE is performing principal component analysis in each local neighborhood, then finding a globally optimal reconstruction of these partial solutions. Although the local embeddings assume linearity, the global solution can preserve non-linearities.
As with non-metric MDS, our implementation of Isomap and LLE scales by the average $d_{ij}$ divided by the average $\hat{d}_{ij}$ to return a final solution.

\section{Performance on Experimental Data}
\label{sec:experiment}

\subsection{Bluetooth measurements for network of 11 devices}
\label{sec:anrg_data}

To evaluate the approaches presented in this work, we considered data collected from 11 devices used in an empty parking lot for which we have recorded locations and RSSI measurements from our previous work on radio frequency localization \cite{yedavalli2005ecolocation}.

\begin{figure}[h!]
    \centering
        \begin{subfigure}[b]{\textwidth}
        \centering
        \includegraphics[trim=0 300 0 0, clip, width=\textwidth]{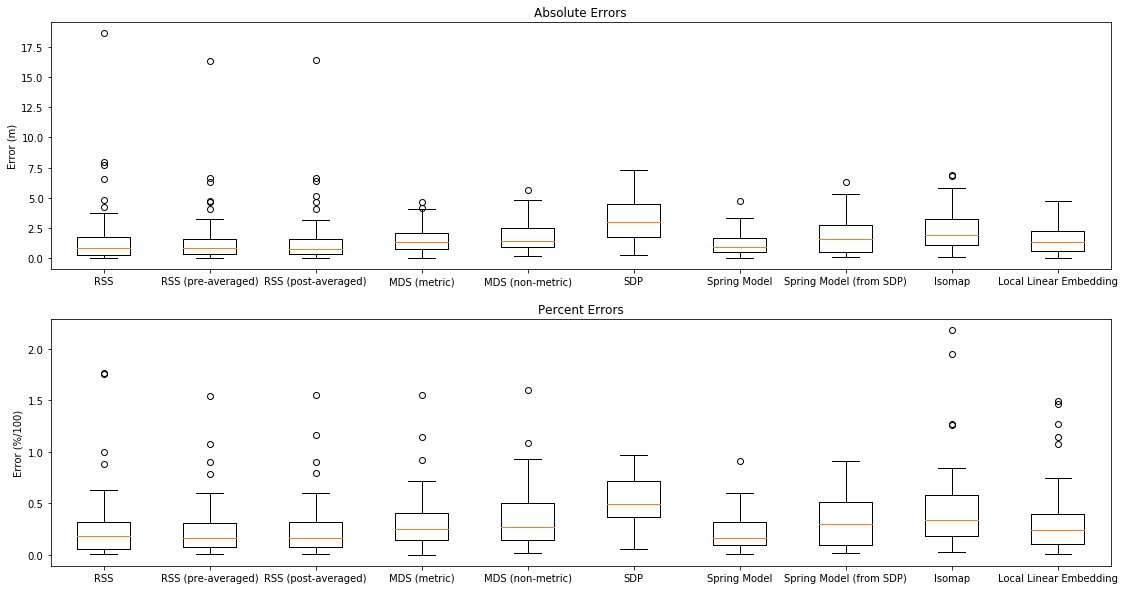}
        \caption{Absolute Errors}
        \end{subfigure}
        \begin{subfigure}[b]{\textwidth}
        \centering
        \includegraphics[trim=0 0 0 300, clip, width=\textwidth]{anrg_results/errors.png}
        \caption{Percent Errors}
        \end{subfigure}
    \caption{Errors on experimental data from 11 node network}
    \label{fig:error_plots}
\end{figure}

Figure \ref{fig:error_plots} shows the spread of errors for each inter-device distance estimate. The absolute error is an intuitive measure of accuracy, but does not consider that large errors may be less significant when estimating large distances. For this reason we show the percent errors as well.
From the plots we see that direct estimation approaches (RSS, pre-averaged, and post-averaged) have outliers, occasionally resulting in high errors. 
SDP, the spring model, and the spring model initialized with SDP all see low maximum percent errors. This is desirable because small variance leads to confident estimates.

\begin{figure}[h!]
    \centering
    \begin{subfigure}[b]{0.45\textwidth}
        \centering
        \includegraphics[trim=0 300 460 20, clip, width=\textwidth]{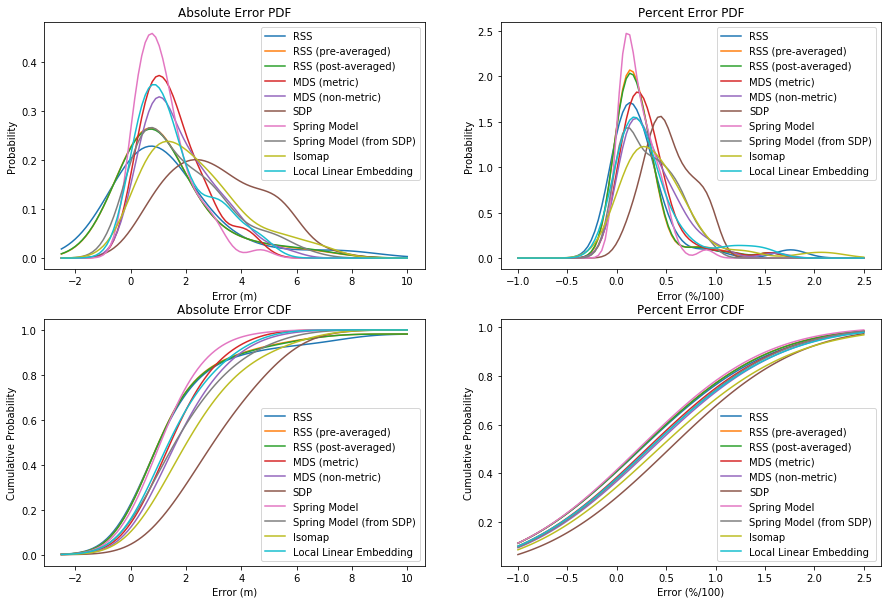}
        \caption{Absolute Error PDF}
    \end{subfigure}
    \begin{subfigure}[b]{0.45\textwidth}
        \centering
        \includegraphics[trim=460 300 0 20, clip, width=\textwidth]{anrg_results/densities.png}
        \caption{Percent Error PDF}
    \end{subfigure}
    \begin{subfigure}[b]{0.45\textwidth}
        \centering
        \includegraphics[trim=0 0 460 320, clip, width=\textwidth]{anrg_results/densities.png}
        \caption{Absolute Error CDF}
    \end{subfigure}
    \begin{subfigure}[b]{0.45\textwidth}
        \centering
        \includegraphics[trim=460 0 0 320, clip, width=\textwidth]{anrg_results/densities.png}
        \caption{Percent Error CDF}
    \end{subfigure}
    \caption{Error density functions}
    \label{fig:densities}
\end{figure}

Fitting probability density function (PDF) and cumulative density function (CDF) curves via gaussian kernel density estimation gives another way to interpret these errors, shown in Figure \ref{fig:densities}. For example, SDP is equally likely to result in an error of 5 meters as an error of 1 meter, with the most likely error of about 2 meters, which can lead to ambiguities. The spring model on the other hand, is unlikely to see errors above 4 meters, and most errors will be around 1 meter. The CDFs allow us to look for stochastic dominance, where the leftmost curve represents the approach which dominates the others. When considering the absolute error no solution is clearly dominant, but when considering the percent error the spring model appears to just barely outperform direct estimation approaches.

Previously we have discussed performance metrics on the accuracy of the results obtained by each approach. It is important to also note discrepancies in complexity of each approach. For a network of $n$ nodes, direct estimation requires worst-case $O(n^2)$ computations. For MDS, the number of computations is bounded by $O(T(n^3+2n^2))$ where $T$ is the number of iterations. $T$ is typically smaller than $n$ and treated as a constant, so the complexity can be thought of as $O(n^3)$. Similarly, the complexity of the spring model is bounded by $O(Tn^2)$ but the number of required iterations has been shown to be approximately $n$, simplifying to $(On^3)$. Semidefinite programming when applied to network localization is on the order of $O(n^3)$ as well. The complexity of Isomap and LLE both depend on the number of neighbors considered, $k$, and are bounded by $O(2nlog(k)log(n))+O(n^2(k+log(n)))+O(2n^2)$ and $O(2nlog(k)log(n)) + O(2nk^3) + O(2n^2)$ respectively. If low-complexity is desirable, direct estimation provides a clear advantage. However under the centralized architecture assumption, all computations can be done at the server.

\begin{figure}[h!]
    \centering
    \includegraphics[trim=0 0 0 20, clip, width=0.76\textwidth]{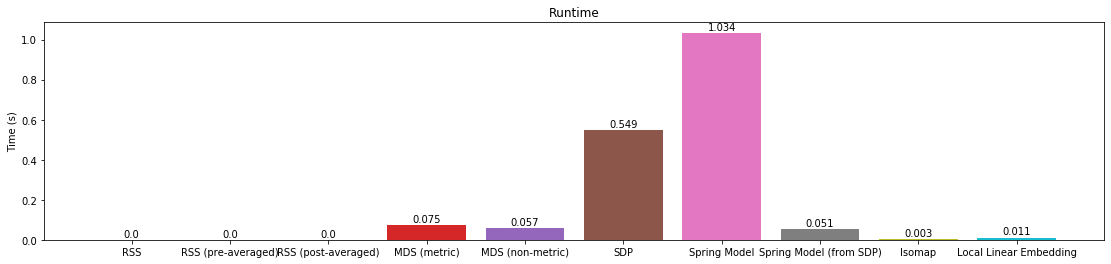}
    \caption{Run times}
    \label{fig:runtimes}
\end{figure}

The results shown in Figure \ref{fig:runtimes} show the run time in seconds of each approach on the experimental data on a personal computer. Note that MDS (metric), MDS (non-metric), and the spring model were each initialized 5 times and only return the results obtained by the best initialization. Increasing the number of initialization can improve performance at the cost of computational expense.


\subsection{Missing RSSI measurements}
\label{sec:missing}
Previously we have noted that not all implementations can handle partial RSSI matrices. Direct estimation and the Scikit-Learn implementation of metric MDS both require estimated distances between all devices and we use a threshold RSSI value representing the noise floor of the reciever for any missing entries, making the assumption that missing RSSI values are a result of large node-node distances. This assumption may be why direct estimation and metric MDS performed well on experimental data.

\begin{figure}[h!]
    \centering
    \includegraphics[trim=0 0 0 662, clip, width=\textwidth]{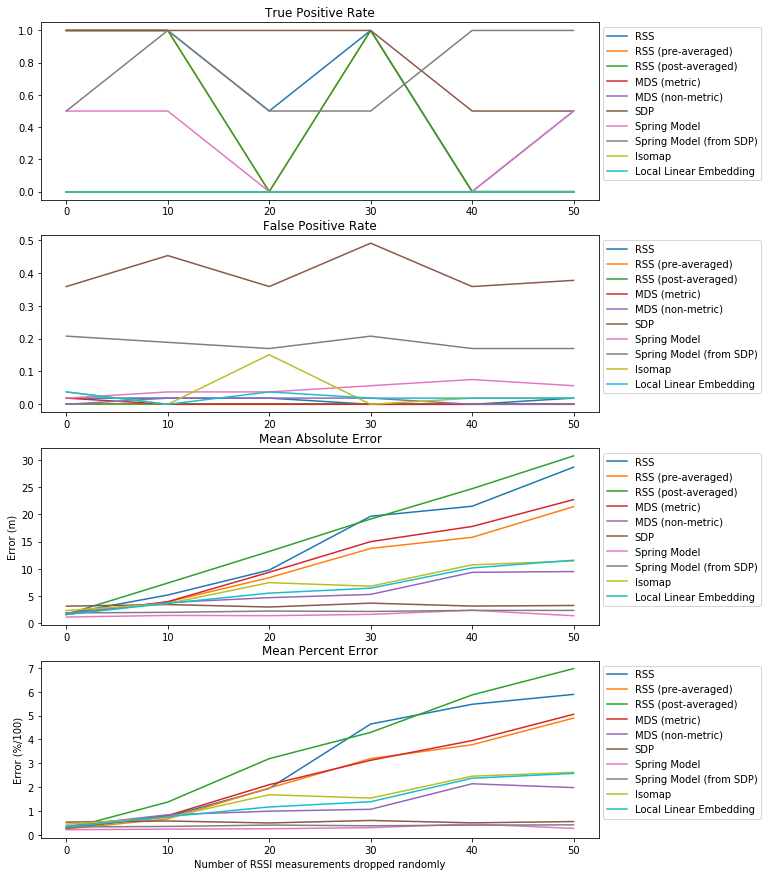}
    \caption{Incomplete RSSI matrices: Mean percent error as RSSI signals are randomly dropped}
    \label{fig:dropped_signals}
\end{figure}

There is ongoing discussion of issues with iPhone-iPhone detection, relating to operating system level restrictions on background app usage \cite{iphone}. Because of this, it seems reasonable to assume that while an Android-Android or iPhone-Android pair within range may record an RSSI measurement, an iPhone-iPhone pair at the same distance may not. This leads us to revisit the assumption that missed measurements are simply a result of long distances.
If we assume instead that some beacon signals are dropped at random, independent of node-node distance, we observe that performance varies significantly across methods. Figure \ref{fig:dropped_signals} shows the change in mean percent error for each approach as the number of dropped RSSI measurements increases, using the same network and measurements as Figure \ref{fig:error_plots}.

We see that SDP, the spring model, and the spring model initialized from SDP are the only approaches which do not see significant performance degradation when RSSI measurements are missing. Isomap, LLE, and non-metric MDS see better performance than direct estimation as well. From this we conclude that unless devices are able to guarantee measurements from all devices within Bluetooth range, network localization algorithms designed for incomplete matrices will see the best results. In fact, leveraging neighboring RSSI measurements through network localization may be the only option for detecting proximity between two iPhones, unless Apple eases app restrictions. 

\subsection{Common contact-logging scenarios}

\begin{figure}[h!]
    \centering
    \includegraphics[trim=0 30 0 0, clip, width=\textwidth]{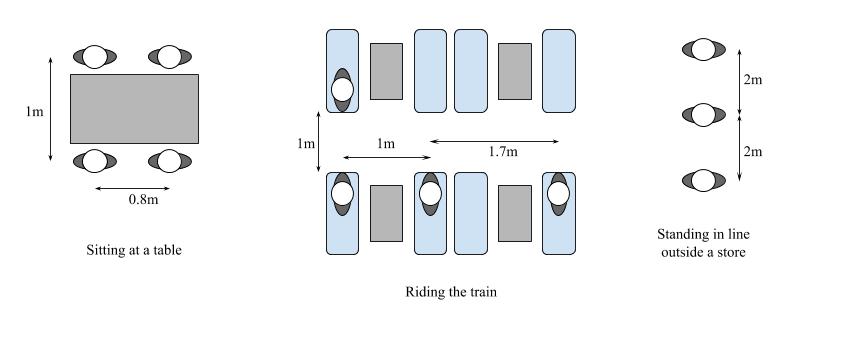}
    \caption{Common scenarios for which data was collected in \cite{leith2020coronavirus}}
    \label{fig:scenario_diagram}
\end{figure}

Figure \ref{fig:scenario_diagram} shows three common scenarios that have been previously considered when analyzing the feasibility of using BLE RSSI measurements for proximity detection \cite{leith2020coronavirus}. The accompanying parameters fitted to data collected in these scenarios are shown in Table \ref{table:trinity_ble_params}. It is immediately clear that the choice of pathloss parameters is highly dependent on the environment. This will make it challenging to design distance estimation algorithms for practical use which are generalizable enough to work indoors, outdoors, on an elevator, etc. For now, we consider each scenario and its accompanying pathloss parameters independently.

\begin{table}[h!]
  \begin{center}
    \caption{Pathloss Parameters fitted to data from \cite{leith2020coronavirus}}
    \label{table:trinity_ble_params}
    \begin{tabular}{l|c|c|c}
      \textbf{Scenario} & Reference power & Pathloss exponent & Noise Standard Deviation \\
       & $P_{0_{dBm}}$ & $\eta$ & $\sigma_{dB}$ \\
      \hline
      Indoors & -62.919 & 2.316 & 3.441 \\
      On a train & -60.452 & 1.364 & 5.054 \\
      Outdoors & -75.014 & 1.7919 & 6.448 \\
    \end{tabular}
  \end{center}
\end{table}

\begin{figure}[h!]
    \centering
    \begin{subfigure}[b]{0.48\textwidth}
        \centering
        \includegraphics[trim=0 260 475 18, clip, width=\textwidth]{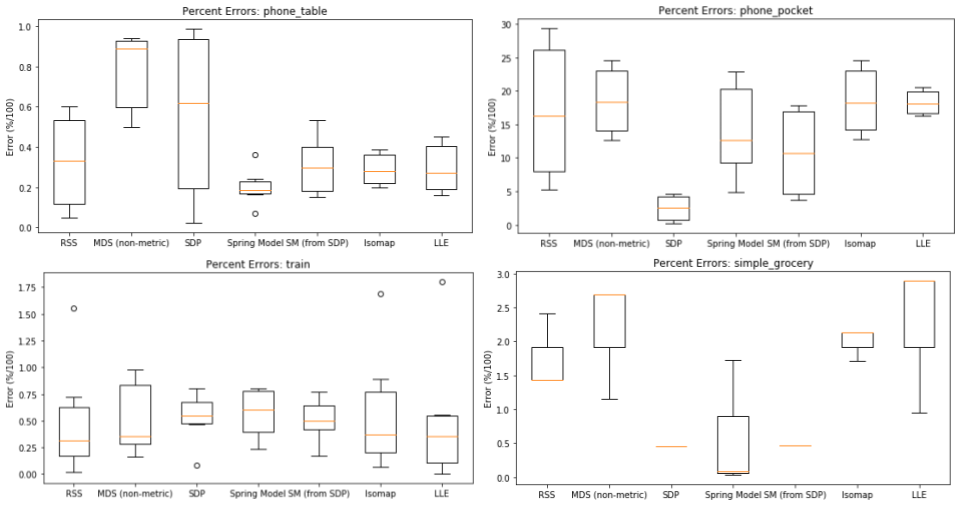}
        \caption{Sitting at a table, phones on table}
        \label{fig:phone_table}
    \end{subfigure}
    \begin{subfigure}[b]{0.48\textwidth}
        \centering
        \includegraphics[trim=480 260 0 18, clip, width=\textwidth]{scenario_results/scenario_results.png}
        \caption{Sitting at a table, phones in pockets (obstruction not detected)}
        \label{fig:phone_pocket}
    \end{subfigure}
    \begin{subfigure}[b]{0.48\textwidth}
        \centering
        \includegraphics[trim=0 0 0 -10, clip, width=\textwidth]{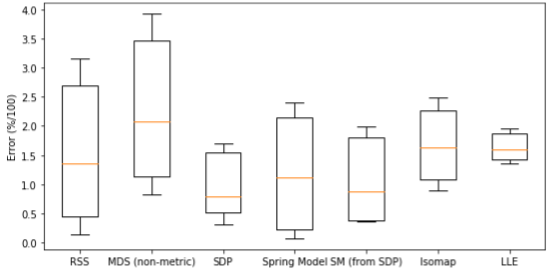}
        \caption{Sitting at a table, phones in pockets (obstruction detected)}
        \label{fig:phone_pocket_predicted}
    \end{subfigure}
    \begin{subfigure}[b]{0.48\textwidth}
        \centering
        \includegraphics[trim=0 0 480 270, clip, width=\textwidth]{scenario_results/scenario_results.png}
        \caption{Riding the train}
        \label{fig:train}
    \end{subfigure}
    \begin{subfigure}[b]{0.48\textwidth}
        \centering
        \includegraphics[trim=475 0 0 270, clip, width=\textwidth]{scenario_results/scenario_results.png}
        \caption{Standing in line outside a store}
        \label{fig:simple_grocery}
    \end{subfigure}
    \caption{Percent errors in common scenarios}
    \label{fig:scenario_results}
\end{figure}

Figure \ref{fig:scenario_results} shows the errors obtained in common scenarios for all approaches which do not require complete RSSI matrices, with direct estimation for comparison. For four people sitting around a table with their phones on the table (\ref{fig:phone_table}), the spring model provides the best performance. However if the four people have their phone inside their pockets (\ref{fig:phone_pocket}), SDP is the only method to find a solution where errors are at most 5x the true distance. In this case the additional attenuation caused by the physical barrier of the pocket, which is not captured in the model, renders all methods unusable. 

It may be possible to use additional sensors on mobile devices like light sensors and proximity sensors to predict whether a phone is in a pocket. If obstructions can be detected, we can compensate for their presence. Figure \ref{fig:phone_pocket_predicted} shows significant improvements in distance estimation when we assume pockets cause 20dB of attenuation \cite{leith2020coronavirus} and their presence is detected.

For four people riding a train (Figure \ref{fig:train}), errors are fairly comparable although non-metric MDS, Spring Model, and Isomap see the best performance with respect to proximity detection (see Table \ref{table:scenario_TPR}). SDP and the SDP-initialized spring model both report 100\% false positive rates, indicating a tendency to underestimate distances. For contact logging, high false positive rates can lead to unnecessary concern and a distrust in the technology.


\begin{table}[h!]
  \begin{center}
    \caption{Proximity detection performance in common scenarios}
    \label{table:scenario_TPR}
    \resizebox{\textwidth}{!}{%
    \begin{tabular}{l|c|c|c|c|c|c|c|c|c|c}
      & \multicolumn{2}{c|}{Phone on table} & \multicolumn{2}{c|}{Phone in pocket} &
      \multicolumn{2}{c|}{Obstruction detected} &
      \multicolumn{2}{c|}{On train} & \multicolumn{2}{c}{Grocery line} \\
      \hline
       & FPR & TPR & FPR & TPR & FPR & TPR & FPR & TPR & FPR & TPR \\
      \hline
      RSS & 0 & 1 & 0 & 0 & 0 & 0.67 & 0.5 & 0.75 & 0 & 0.5 \\
      MDS (non-metric) & 0 & 1 & 0 & 0 & 0 & 0.33 & 0 & 0.75 & 0 & 0.5 \\
      SDP & 0 & 1 & 0 & 0.33 & 0 & 0.5 & 1 & 1 & 0 & 0.5 \\
      Spring Model & 0 & 1 & 0 & 0 & 0 & 0.5 & 0 & 0.75 & 0 & 0.5 \\
      Spring Model (from SDP) & 0 & 1 & 0 & 0 & 0 & 0.5 & 1 & 1 & 0 & 0.5 \\
      Isomap & 0 & 1 & 0 & 0 & 0 & 0.33 & 0 & 0.75 & 0 & 0.5 \\
      LLE & 0 & 1 & 0 & 0 & 0 & 0 & 0 & 0.5 & 0 & 0.5 \\
    \end{tabular}
    }
  \end{center}
\end{table}


Note that for phones in pockets, only a third of device pairs which were closer than 2 meters were detected by SDP, and none were detected by the other approaches. 
However when the pocket obstruction is detected and accounted for, up to two thirds of true positives were detection by direct estimation.
From this we conclude that while network localization algorithms can improve inter-device distance estimation, they cannot alone overcome the fact that patterns in received signal strength depend heavily on the environment and obstructions. Automatically detecting environmental factors presents an area for future research.

\section{Performance on Simulated Data}
\label{sec:simulated}

To test our methods on scenarios which were not already captured in available
real-world datasets we developed techniques for simulating networks while
attempting to preserve the noise profile seen in real-world experiments. This
was used to test the viability of approaches as network size scales, how well
different approaches handled varying levels of sparsity in the RSSI matrix, and
the effects of different forms of real-world noise on the performance of the
different distance estimation approaches. Such synthetic situations are useful
as they allow for modeling of a diverse set of scenarios while having a
ground-truth solution readily available.
These synthetic scenarios leveraged the data captured in
\cite{leith2020coronavirus}. 

\subsection{Synthetic Network Generation}
\label{sec:synthetic}

We randomly generated networks of device locations
and determined inter-device RSSI based upon pair-wise device distance. If the
distance was exactly a distance for which RSSI was measured in
\cite{leith2020coronavirus}, the RSSI was randomly sampled from one of the
reported RSSIs at the given distance from the data. If the distance was between
two measured distances from the data, RSSI was sampled from both respective
distances and linear interpolation was performed between the two measurements.
If the distance was below any measured distance, RSSI was sampled from the
smallest distance measured (0.5 m) as RSSI was found to change little as that
distance continued to shrink. If the simulated distance was above the maximum
measured distance (4 m), the devices were considered to be out of range of each
other.

For all scenarios we assumed a ten percent chance that a given
measurement would be missed. For all results shown, excluding Table
\ref{table:summary_w_pockets_no_misses}, the RSSI data generated came from
measurements between two Google Pixel phones in an outdoor environment.

\subsection{Summary of Results}
\label{sec:summary}


In Tables \ref{table:summary_pixel_10_misses} and
\ref{table:summary_w_pockets_no_misses} we present the average values of our
estimation approaches over a wide range of simulated network sizes and
densities. Table \ref{table:summary_pixel_10_misses} represents
scenarios where measurements were sampled from Google Pixels in open air, such as in one's hand or on a table. In comparison, Table \ref{table:summary_w_pockets_no_misses} represents scenarios where measurements were sampled from varying devices in varying contexts, including in one's pocket or purse.



\begin{table}[h!]
  \caption{Summarized Results for Single Device Type in Open Air}
  \label{table:summary_pixel_10_misses}
  \resizebox{\textwidth}{!}{%
    \begin{tabular}{l|r|r|r|r|r|r|r|r|r|r}
      {} &  Isomap &    LLE &  MDS Metric &  MDS Non-Metric &  RSS &  RSS Post &  RSS Pre &    SDP &  SDP + Spring &  Spring Model \\
      \hline

      Avg \% Error &    0.50 &   0.50 &        8.10 &            0.50 &      6.60 &               6.60 &              6.60 &    0.50 &             0.40 &                 0.30 \\
      Max \% Error &   10.50 &  10.40 &      103.50 &           26.80 &    307.30 &             307.30 &            307.30 &    3.20 &             3.50 &                11.10 \\
      TPR         &    0.68 &   0.65 &        0.06 &            0.62 &      0.84 &               0.84 &              0.84 &    0.98 &             0.94 &                 0.86 \\
      FPR         &    0.16 &   0.16 &        0.01 &            0.21 &      0.10 &               0.09 &              0.09 &    0.51 &             0.34 &                 0.15 \\
    \end{tabular}
  }
\end{table}

\begin{table}[h!]
  \caption{Summarized Results for Varying Device Types in Varying Context}
  \label{table:summary_w_pockets_no_misses}
  \resizebox{\textwidth}{!}{%
    \begin{tabular}{l|r|r|r|r|r|r|r|r|r|r}
      {} &  Isomap &    LLE &  MDS Metric &  MDS Non-Metric &  RSS &  RSS Post &  RSS Pre &    SDP &  SDP + Spring &  Spring Model \\
      \hline
Avg \% Error &    1.30 &   1.20 &        8.80 &            1.30 &      7.60 &               7.60 &              7.30 &   0.80 &             0.70 &                 0.90 \\
Max \% Error &   19.30 &  19.90 &      102.10 &           53.70 &    261.30 &             261.90 &            261.30 &  22.30 &            20.40 &                27.40 \\
TPR         &    0.36 &   0.34 &        0.03 &            0.18 &      0.39 &               0.29 &              0.33 &   0.62 &             0.63 &                 0.40 \\
FPR         &    0.07 &   0.08 &        0.00 &            0.08 &      0.08 &               0.04 &              0.04 &   0.45 &             0.42 &                 0.12 \\

\end{tabular}
  }
\end{table}

As can be seen, the methods used showed substantial decrease in performance when
the pathloss parameters greatly varied and were not well understood.
These results indicate that a great challenge in deploying these techniques is
the ability to accurately understand or control the testing environment.
Investigation of the substantial drop-off in performance showed in Table
\ref{table:summary_w_pockets_no_misses} indicates the change in performance is
largely due to the substantial increase in attenuation seen in certain
configurations. The attenuation from a pocket or purse is commonly in the range
of 15 to 20 dB \cite{leith2020coronavirus}. The major takeaway from this result
is that the successful deployment of these approaches depends on being able to
sense the environment and accurately predict signal attenuation across a wide
range of scenarios. With that being said, we note that the spring model initialized with SDP achieves the best distance estimation performance and the best proximity detection if false negatives and false positives are equally undesirable.

\subsection{Dense Networks}
\label{sec:dense}

\begin{figure}[h!]
    \centering
    \includegraphics[width=\textwidth]{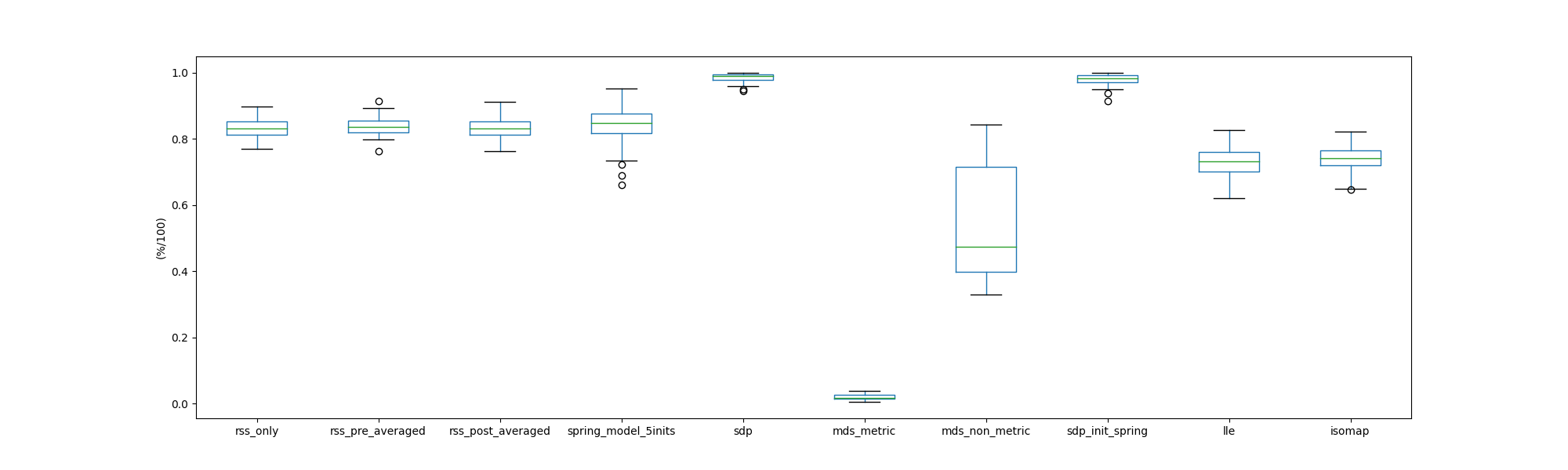}
    \caption{True Positive Rates for 50 devices in 100 square meters}
    \label{fig:small_dense_tp}
\end{figure}

\begin{figure}[h!]
    \centering
    \includegraphics[width=\textwidth]{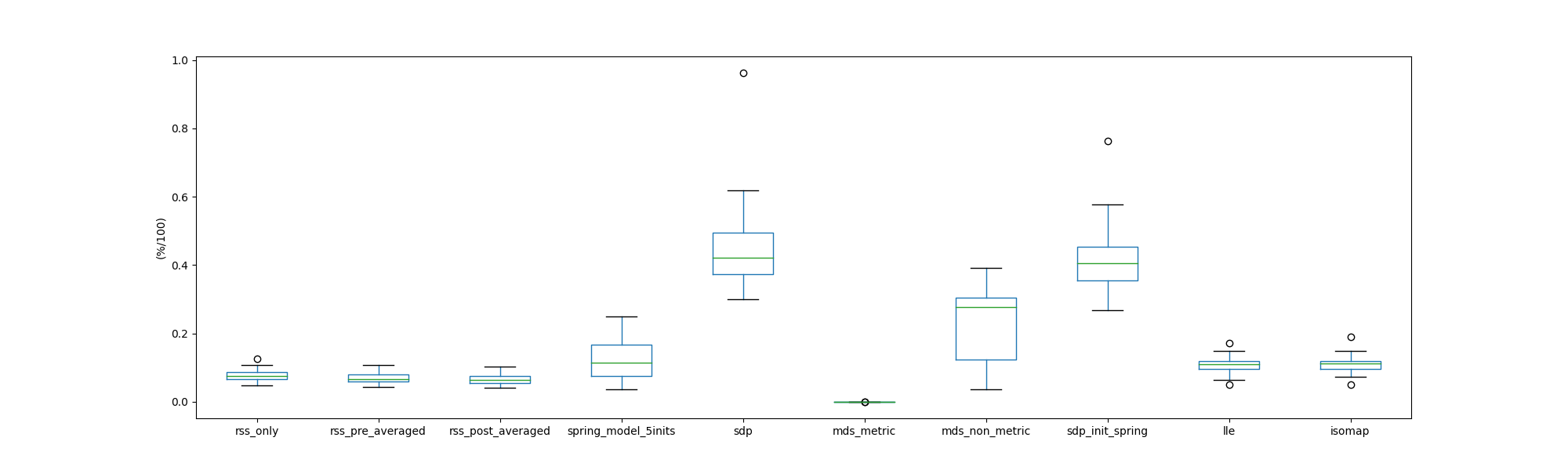}
    \caption{False Positive Rates for 50 devices in 100 square meters}
    \label{fig:small_dense_fp}
\end{figure}

\begin{figure}[h!]
    \centering
    \includegraphics[width=\textwidth]{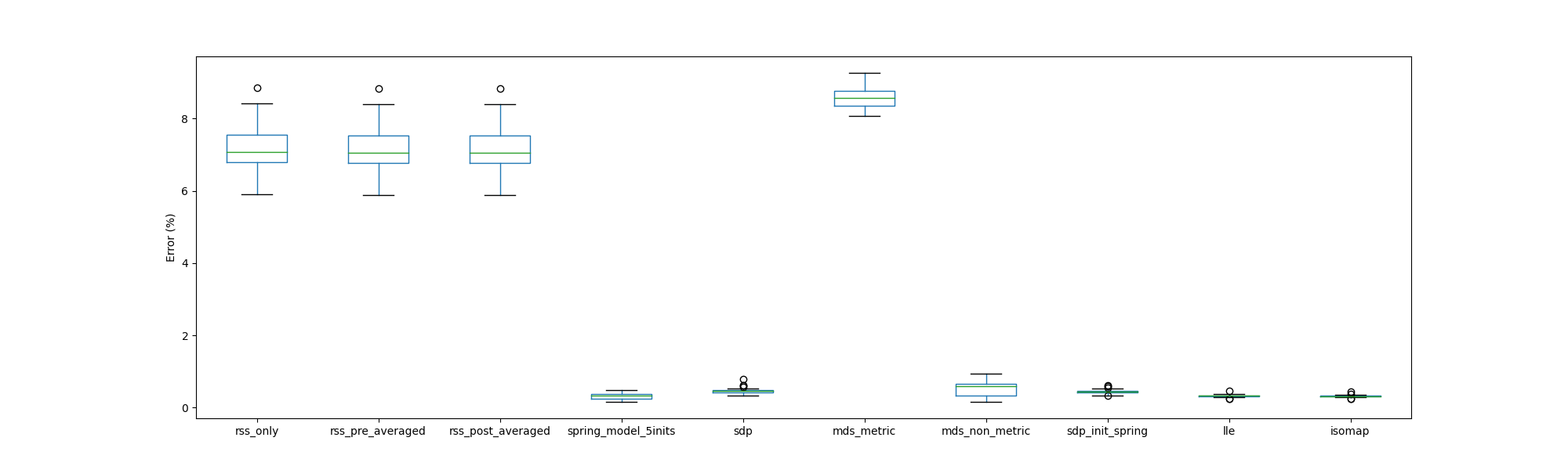}
    \caption{Mean Percent Error for 50 devices in 100 square meters}
    \label{fig:small_dense_mpe}
\end{figure}

We simulated 50 scenarios where 50 devices were randomly positioned across a 100
square meter field. Figures \ref{fig:small_dense_tp} and
\ref{fig:small_dense_fp} show the true positive rate (TPR) and false positive
rate (FPR) achieved across the simulations. Approaches using semidefinite
programming (SDP, SDP + Spring) appear to generally outperform the others in
terms of TPR while also generally under-performing all others in terms of FPR,
indicating they tend to return positive predictions. 
Direct estimation approaches (RSS, pre-averaged RSS, and post-averaged RSS) lead to high true positive rates and low false positive rates, indicating good performance with respect to proximity detection. However examining Figure 
\ref{fig:small_dense_mpe}, we see that methods which allow for missing
measurements greatly outperform the direct approaches with respect to distance estimation.
The spring model, Isomap, and LLE appear well-suited for both proximity detection and distance estimation.

\begin{figure}[h!]
    \centering
    \includegraphics[width=\textwidth]{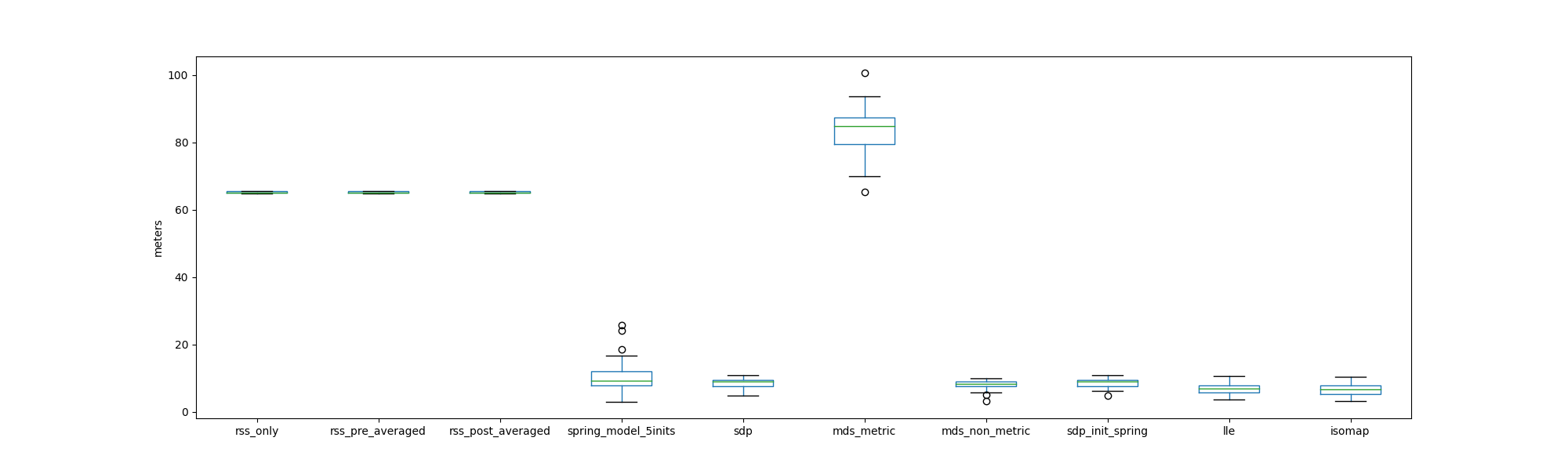}
    \caption{Maximum Error for 50 devices in 100 square meters}
    \label{fig:small_dense_max}
\end{figure}

\subsection{Sparse Networks}
\label{sec:sparse}

The previous simulated data resulted in dense network graphs. We are also
interested in performance in sparse networks where devices are spread across
large areas therefore RSSI measurements are not available for many device pairs.
We simulate 50 devices randomly positioned across a 400 square meter
field. We observe similar patterns, 
with SDP-based methods leading to the most true positives but also the most false positives, and direct methods leading to good proximity detection accuracy but poor distance estimate accuracy.
These results are omitted for
brevity. 

\subsection{Large Networks}
\label{sec:large}

\begin{figure}[h!]
    \centering
    \includegraphics[width=\textwidth]{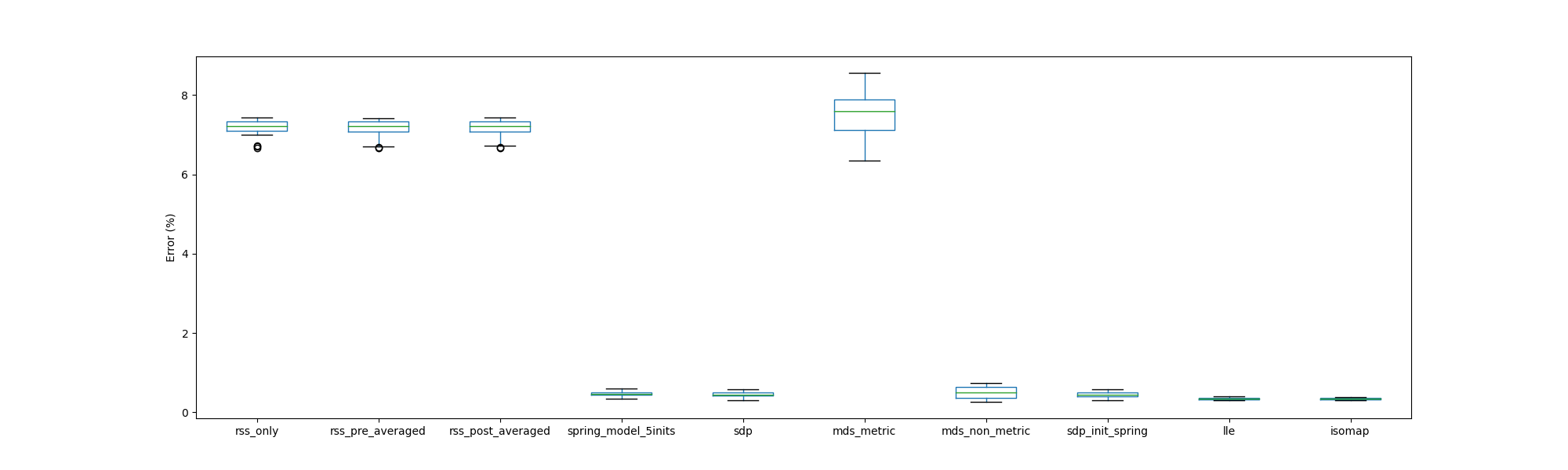}
    \caption{Mean Percent Error for 100 nodes in 400 square meters}
    \label{fig:large_mpe}
\end{figure}

We are also interested in examining how these approaches scale to larger
networks. For this we simulate 100 nodes over a 400 square meter field. For
these large simulated networks (Figure \ref{fig:large_mpe}), the proximity detection and distance estimation results agree with those obtained in Sections
\ref{sec:dense} and \ref{sec:sparse}. 
As discussed in section \ref{sec:experiment}, runtimes increased across all methods as the number of nodes grew. We noted that the open-source optimized implementations of network localization algorithms saw average runtimes of 2-5 seconds for 100 node networks on our hardware, likely acceptable for most use cases. However
if scalably fast
calculations are the main priority, the averaged direct estimation approaches
are the best choice.   

\section{Conclusions and Future Work}
\label{sec:conclusion}

In this work we have surveyed several approaches to estimating inter-device distances using network localization algorithms to explore the emerging application of automated contact logging. 
We have found that the choice of algorithm is quite nuanced, and the success of algorithms is largely dependent on how  representative the pathloss model is of the actual environment. 
Our experiments indicate that in ideal settings when all RSSI measurements are available, direct estimation provides the best proximity detection and lowest complexity while the spring model narrowly outperforms direct approaches with respect to inter-device distance estimation.
Under more realistic assumptions, SDP sees good distance estimate accuracy and is the most likely to identify all true positives (distances less than 2 meters), but it is also the most likely to lead to false positives. The spring model again sees the best distance estimation and good proximity detection, but there is some variance in performance as the result depends on a random initialization.

\begin{figure}
    \centering
    \includegraphics[width=\textwidth]{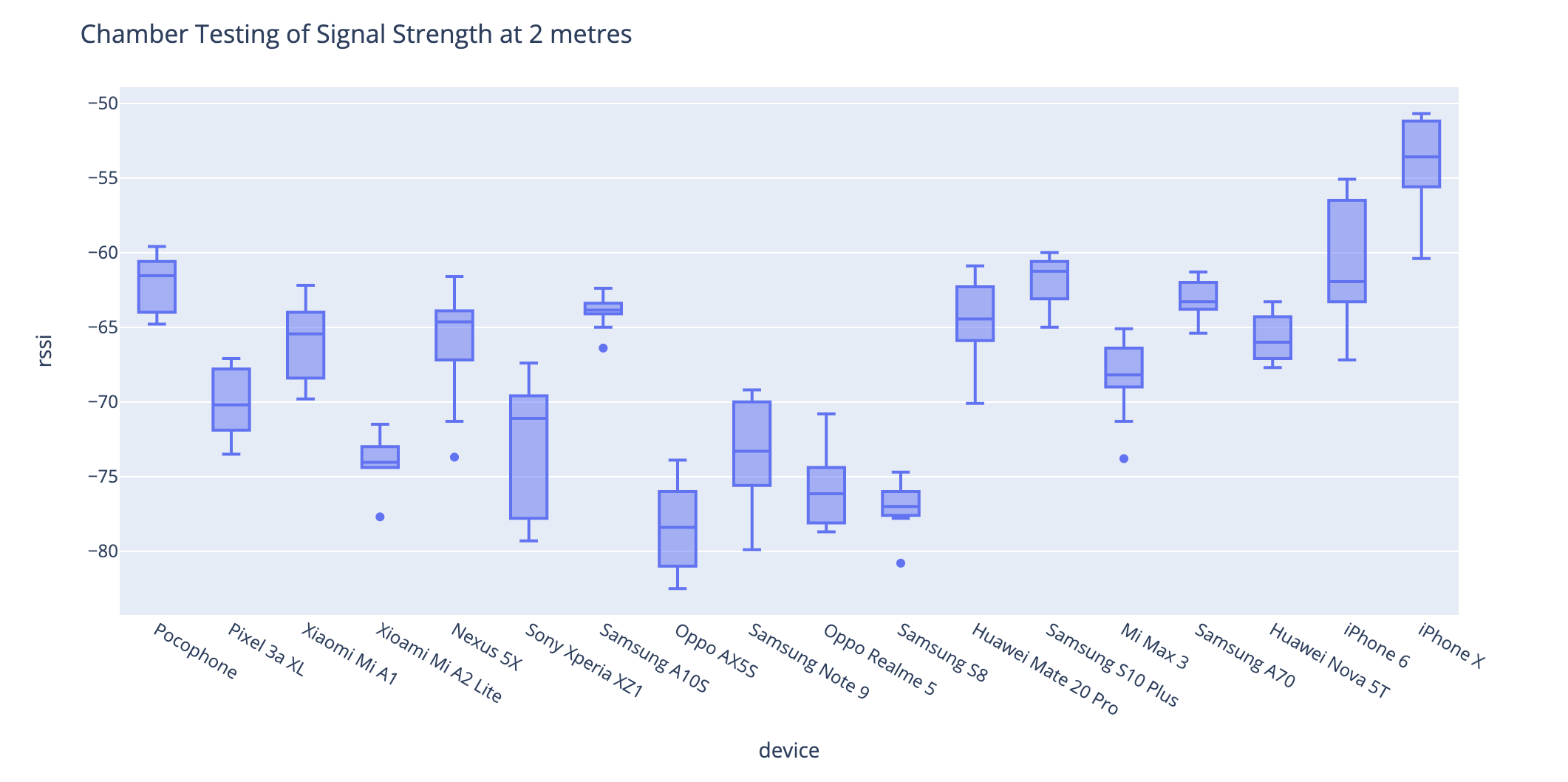}
    \caption{RSSI data for devices at 2 meters available from https://github.com/opentrace-community \cite{bay2020bluetrace}}
    \label{fig:device_models}
\end{figure}

Importantly, we have also observed that context (i.e. indoors, outdoors, in a pocket, etc) has a greater effect on the resulting distance estimate matrix than the network localization approach. Similarly, the relationship between distance and signal strength is not the same for all devices. The OpenTrace Community has provided access to the RSSI measurements which correspond to 2 meters for 18 phone models in Figure \ref{fig:device_models} \cite{bay2020bluetrace}. 
With the increasing availability of this data, we can specialize our RSSI-to-distance model based on device type. We have shown that the ability to gauge context via data from ambient light sensors and/or proximity sensors to detect obstructions would increase performance across all approaches.

All of the methods presented in this work are anchorless, ``one-shot" localization approaches, meaning we assume no locations are known in advance, all measurements are taken at one time, and no data is available before or after this moment. Further improvements could be made taking into account the location of cell phone towers or other potential beacons, or by leveraging time sequences of measurements and user mobility models.

As accurately estimating distance from Bluetooth measurements is a challenging and timely problem, we intend to continue research on improving these models and algorithms. We are working closely with the Covid Community Alert team, and plan on incorporating the findings from this work in their open-source anonymous protocol.


\bibliographystyle{elsarticle-num}

\bibliography{ref}

\end{document}